\documentclass[twocolumn,tighten,times]{aastex61}

\newcommand{\kmsmpc}{km\,s$^{-1}$\,Mpc$^{-1}$}
\newcommand{\kms}{km\,s$^{-1}$}

\newcommand{\hst}{\textit{HST}}

\newcommand{\mbar}{$\overline m$}
\newcommand{\Mbar}{$\overline M$}

\newcommand{\mM}{\ensuremath{(m{-}M)}}
\newcommand{\colorbar}{$(\overline m_{110}\,{-}\,\overline m_{160})$}
\newcommand\zacs{\ifmmode z_{850}\else$z_{850}$\fi}
\newcommand\gacs{\ifmmode g_{475}\else$g_{475}$\fi}
\newcommand\jwfc{\ifmmode J_{110}\else$J_{110}$\fi}
\newcommand\hwfc{\ifmmode H_{160}\else$H_{160}$\fi}

\newcommand\gz{\ensuremath{(g_{475}{-}z_{850})}}
\newcommand\JH{\ensuremath{(J_{110}{-}H_{160})}}

\newcommand\sersic{S\'ersic}

\received{}
\revised{}
\accepted{}

\shorttitle{A Precise Distance to NGC\,4993}
\shortauthors{Cantiello, Jensen, Blakeslee et al.}

\begin{document}

\title{A Precise Distance to the Host Galaxy of the Binary Neutron Star Merger GW170817\\ Using Surface Brightness Fluctuations\footnote{Based on observations with the NASA/ESA \textit{Hubble
Space Telescope}, obtained at the Space Telescope Science Institute, 
which is operated by the Association of Universities for Research in 
Astronomy, Inc., under NASA contract NAS 5-26555. 
These observations are associated with Program \#15329 (PI E.~Berger, 
\#14771 (PI N.~Tanvir), and \#14804 (PI A.~Levan).}}

\correspondingauthor{Michele Cantiello}

\author[0000-0003-2072-384X]{Michele Cantiello}
\affil{INAF--Osservatorio Astronomico d'Abruzzo, Via M. Maggini snc, 64100, Teramo, Italy}
\email{\tt cantiello@oa-abruzzo.inaf.it}

\author[0000-0001-8762-8906]{J.\ B.\ Jensen}
\affil{Utah Valley University, Orem, UT 84058, USA}

\author[0000-0002-5213-3548]{J.\ P.\ Blakeslee}
\affil{NRC Herzberg Astronomy \& Astrophysics Research Centre, Victoria, BC, Canada}
\affil{Gemini Observatory, Casilla 603, La Serena 1700000, Chile}

\author[0000-0002-9392-9681]{E.\ Berger}
\affil{Harvard-Smithsonian Center for Astrophysics, 60 Garden Street, Cambridge, MA 02138, USA}

\author[0000-0001-7821-9369]{A.\ J.\ Levan}
\affil{Department of Physics, University of Warwick, Coventry CV4 7AL, UK}

\author[0000-0003-3274-6336]{N.\ R.\ Tanvir}
\affil{Department of Physics and Astronomy, University of Leicester, Leicester LE1 7RH, UK}

\author[0000-0002-5577-7023]{G.\ Raimondo}
\affil{INAF--Osservatorio Astronomico d'Abruzzo, Via M. Maggini snc, 64100, Teramo, Italy}

\author{E.\ Brocato}
\affil{INAF--Osservatorio Astronomico di Roma, Via di Frascati 33, 00078, Monteporzio Catone,  Italy}

\author{K.~D.~Alexander}
\affil{Harvard-Smithsonian Center for Astrophysics, 60 Garden Street, Cambridge, MA 02138, USA}

\author{P.~K.~Blanchard}
\affil{Harvard-Smithsonian Center for Astrophysics, 60 Garden Street, Cambridge, MA 02138, USA}

\author{M.~Branchesi}
\affil{Gran Sasso Science Institute (GSSI), Viale Francesco Crispi 7, 67100, L'Aquila, Italy}
\affil{INFN--Laboratori Nazionali del Gran Sasso, Via G. Acitelli, 22, 67100, Assergi, Italy}

\author{Z.~Cano}
\affil{Instituto de Astrof\'isica de Andaluc\'ia (IAA-CSIC), Glorieta de la Astronom\'ia s/n, 18008 Granada, Spain}

\author{R.~Chornock}
\affil{Astrophysical Institute, Department of Physics and Astronomy, 251B Clippinger Lab, Ohio University, Athens, OH 45701, USA}

\author{S.~Covino}
\affil{INAF--Osservatorio Astronomico di Brera, Via Bianchi 46, 23807, Merate, Italy}

\author{P.~S.~Cowperthwaite}
\affil{Harvard-Smithsonian Center for Astrophysics, 60 Garden Street, Cambridge, MA 02138, USA}

\author{P.\ D'Avanzo}
\affil{INAF--Osservatorio Astronomico di Brera, Via Bianchi 46, 23807, Merate, Italy}

\author{T.~Eftekhari}
\affil{Harvard-Smithsonian Center for Astrophysics, 60 Garden Street, Cambridge, MA 02138, USA}

\author{W.~Fong}
\affil{Center for Interdisciplinary Exploration and Research in Astrophysics (CIERA) and Department of Physics and Astronomy, Northwestern University, Evanston, IL 60208}

\author{A.~S.~Fruchter}
\affil{Space Telescope Science Institute, 3700 San Martin Drive, Baltimore, MD 21218, USA}

\author{A.~Grado}
\affil{INAF--Osservatorio Astronomico di Capodimonte, salita Moiariello 16, 80131, Napoli, Italy}

\author{J.~Hjorth}
\affil{Dark Cosmology Centre, Niels Bohr Institute, University of Copenhagen, Juliane Maries Vej 30, DK-2100 Copenhagen, Denmark}

\author{D.\ E.\ Holz}
\affil{Enrico Fermi Institute, Department of Physics, Department of Astronomy and Astrophysics and Kavli Institute for Cosmological Physics, University of Chicago, Chicago, IL 60637, USA}

\author{ J.\ D.\ Lyman}
\affil{Department of Physics, University of Warwick, Coventry CV4 7AL, UK}

\author{I.~Mandel}
\affil{Institute of Gravitational Wave Astronomy and School of Physics and Astronomy, University of Birmingham, Birmingham, B15 2TT, UK}

\author{R.~Margutti}
\affil{Center for Interdisciplinary Exploration and Research in Astrophysics (CIERA) and Department of Physics and Astronomy, Northwestern University, Evanston, IL 60208}

\author{M.~Nicholl}
\affil{Harvard-Smithsonian Center for Astrophysics, 60 Garden Street, Cambridge, MA 02138, USA}

\author{V.~A.~Villar}
\affil{Harvard-Smithsonian Center for Astrophysics, 60 Garden Street, Cambridge, MA 02138, USA}

\author{P.~K.~G.~Williams}
\affil{Harvard-Smithsonian Center for Astrophysics, 60 Garden Street, Cambridge, MA 02138, USA}

\begin{abstract}
The joint detection of gravitational waves and electromagnetic
radiation from the binary neutron star (BNS) merger GW170817 has
provided unprecedented insight into a wide range of physical
processes: heavy element synthesis via the $r$-process; the production
of relativistic ejecta; the equation of state of neutron stars and the
nature of the merger remnant; the binary coalescence timescale; and a
measurement of the Hubble constant via the ``standard siren''
technique.  In detail, all of these results depend on the distance to
the host galaxy of the merger event, NGC\,4993.  In this paper we
measure the surface brightness fluctuation (SBF) distance to NGC\,4993
in the F110W and F160W passbands of the Wide Field Camera~3 Infrared
Channel on the \emph{Hubble Space Telescope} (\hst).  For the
preferred F110W passband we derive a distance modulus of
$\mM=33.05\pm0.08\pm0.10$ mag, or a linear distance
$d=40.7\pm1.4\pm1.9$ Mpc (random and systematic errors, respectively);
a virtually identical result is obtained from the F160W data. This is
the most precise distance to NGC\,4993 available to date.  Combining
our distance measurement with the corrected recession velocity of
NGC\,4993 implies a Hubble constant $H_0=71.9\pm 7.1$ \kmsmpc.  A
comparison of our result to the GW-inferred value of $H_0$ indicates a
binary orbital inclination of $i\,{\gtrsim}\,137~\deg$. The SBF
technique can be applied to early-type host galaxies of BNS mergers to
${\sim\,}100$~Mpc with \hst\ and possibly as far as ${\sim\,}300$~Mpc
with the {\it James Webb Space Telescope}, thereby helping to break
the inherent distance-inclination degeneracy of the GW data at
distances where many future BNS mergers are likely to be detected.
\end{abstract}


\section{Introduction} 
\label{sec:intro}

On 2017 August 17, the Advanced LIGO and Virgo gravitational wave (GW)
observatories detected a binary neutron star (BNS) merger for the
first time \citep[GW170817,][]{abbott17prl}. The merger was followed
about 1.7 s later by a short-duration gamma-ray burst, detected by
{\it Fermi} and INTEGRAL
\citep[GRB\,170817A,][]{abbott17grb,savchenko17}.  Optical and
near-infrared (NIR) follow-up observations of the GW localization
region led to the identification of a counterpart in the galaxy
NGC\,4993
\citep{abbott17grb,arcavi17,coulter17,lipunov17,soaressantos17,valenti17}.
Subsequent photometric and spectroscopic observations in the
ultraviolet, optical, and NIR revealed the signatures of a
``kilonova'', a transient powered by the radioactive decay of
$r$-process material synthesized in the merger ejecta (e.g.,
\citealt{cowperthwaite17,chornock17,nicholl17,pian17,smartt17,tanvir17,villar17}).
Rising X-ray and radio emission produced by a separate relativistic
ejecta component were detected with a delay of about two weeks
\citep{alexander17,haggard17,hallinan17,kim17,margutti17,mooley17,troja17}.
In addition, studies of NGC\,4993 itself have established it to be an
early-type galaxy dominated by an evolved stellar population with a
median age of ${\sim}10$ Gyr, and negligible present-day star
formation activity \citep{blanchard17,im17,levan17}.  Finally,
combining the redshift of NGC\,4993 with the distance measured from
the GW data, Hubble constant values of $H_0\,{=}\,70^{+12}_{-8}$
\kmsmpc\ \citep{abbott17siren} and $H_0\,{=}\,75^{+12}_{-10}$
\kmsmpc\ \citep{guidorzi17} were estimated.  The large uncertainties
in these measurements are dominated by the distance-inclination
degeneracy inherent in the GW signal.

In detail, all of these transformative results depend on the distance
to NGC\,4993, which has been presently measured in two ways. First,
from the GW signal itself, using the exact sky location available from
the EM counterpart, the distance is estimated to be
$d\,{=}\,43.8^{+2.9}_{-6.9}$ Mpc \citep{abbott17siren}; the
uncertainty is dominated by a fundamental degeneracy with the
inclination of the binary's orbit relative to the plane of the sky.
Second, using the Fundamental Plane (FP) relation the distance is
estimated to be $d\,{=}\,44.0\,{\pm}\,7.5$ Mpc \citep{hjorth17} or
$d\,{=}\,37.7\,{\pm}\,8.7$ Mpc \citep{im17}; the ${\sim}20\%$
uncertainties and difference between the two FP estimates is typical
for this method when applied to individual galaxies
\citep[e.g.,][]{blake02}. \citet{hjorth17} also evaluated a distance
of $d\,{=}\,40.4\pm3.4$ Mpc to NGC\,4993, from the galaxy redshift and
adopting the value for $H_0$ from \citet{riess16}; by combining the FP
and the $H_0$-dependent distance the authors obtained
$d\,{=}\,41.0\pm3.1$ Mpc.

Since NGC\,4993 is an early-type galaxy and too distant for individual
stars to be resolved, yet near enough that peculiar velocities
typically exceed 10\% of the Hubble velocity, the options for a
high-quality distance are quite limited.  Of the six high-precision
distance-determination methods discussed in the comprehensive review
by \citet[][]{freedman10}, three (Cepheids, tip of the red giant
branch, and Tully-Fisher) are either impractical or impossible.  Two
other methods are presently impossible because no water masers or Type
Ia supernovae have been observed in NGC\,4993 to date.  This leaves
surface brightness fluctuations (SBF) as the only viable
high-precision method for determining the distance.  When applied with
modern wide-field instruments on the \emph{Hubble Space Telescope}
(\hst), the SBF method has an intrinsic scatter of ${\lesssim}\,5$\%
\citep{blake09,blake13,jensen15}, and indeed it has already been
proposed for determining the distance to NGC\,4993 with high precision
\citep{hjorth17}. Here we use \hst\ observations collected as part of
the follow-up observations of GW170817 to measure an SBF distance to
NGC\,4993.  Our analysis results in the most precise distance
available to date.

\section{Observations and Data Processing
\label{sec:obs}}

Thanks to the combination of high angular resolution, stable image
quality, and low background, accurate SBF measurements can be made for
any bright early-type galaxy within ${\sim}\,80$ Mpc in only a single
orbit with one of the wide passband filters of the Wide Field Camera 3
Infrared Channel (WFC3/IR) on \hst\ \citep{jensen15}.  To achieve the
best precision and to avoid systematic errors, we processed and
analyzed \hst\ imaging data from three different WFC3/IR programs that
targeted NGC\,4993 as part of the follow-up of GW170817.  All three
programs (GO-15329, PI E.~Berger; GO-14804, PI A.~Levan; GO-14771,
PI N.~Tanvir) collected data in the F110W and F160W filters
(\jwfc\ and \hwfc\, hereafter), both of which have been previously
calibrated for the SBF method \citep{jensen15}. We also used data in
the F475W and F850LP filters (hereafter $g_{475}$ and $z_{850}$) of
the Advanced Camera for Surveys (ACS) from GO-15329 to derive the
galaxy \gz\ color for calibrating the absolute SBF magnitude. The data
from GO-15329 were sufficiently deep ($1102$ s in each filter) for
measuring SBF on their own; the data from the two other programs were
combined to achieve the required depth ($893$ s total in each
filter). The two resulting data sets were processed and analyzed
independently as described below in Section~\ref{sec:sbfmeasurements}.

We reprocessed the raw \jwfc\ and \hwfc\ WFC3/IR images from the
Mikulski Archive for Space Telescopes before proceeding with the SBF
analysis. There are two reasons for this. First, the SBF analysis is
performed using the spatial power spectrum of the Fourier-transformed
image.  When images are geometrically corrected and combined using
pixel interpolation algorithms (as is the default in the WFC3
pipeline), correlations are introduced in the noise of neighboring
pixels, which can adversely affect the SBF fitting procedure
\citep[e.g.,][]{cantiello05,mei05iv}. Thus, we used only integer pixel
shifts when combining exposures without correcting for geometrical
distortion; this ensured that the power spectrum of the noise in the
resulting stacked image was flat (white noise), as desired. Due to the
spatial distortion of the WFC3/IR, the final stacked images have plate
scales that differ in $x$ and $y$ by 10\%, but this does not affect
the SBF analysis, as long as the template point spread function (PSF)
shares the same distortion. For galaxies with significant color
gradients, it also requires that the color map be transformed in a
consistent way before determining the colors \citep{jensen15}.

The second reason for reprocessing the raw WFC3/IR exposures is to
identify and correct the ones affected by the diffuse He emission at
1.083 $\mu$m, generated by metastable helium atoms in the Earth's
upper atmosphere, which causes a variable background level in the
\jwfc\ filter \citep{brammer14}.  We processed all of the raw IR
images using a routine written by
G.~Brammer\footnote{https://github.com/gbrammer/wfc3} that searches
for varying rates of flux accumulation in a WFC3/IR MULTIACCUM
sequence and corrects for the variations by fitting a linear trend to
the background level as a function of time. After the background was
linearized for a MULTIACCUM exposure sequence, the WFC3 calibration
pipeline was used again on each exposure to regenerate the processed
images \citep[see][for further discussion]{goullaud18}.  These images
were then registered and combined using integer pixel offsets, as
discussed above.

We corrected all photometric measurements for Galactic extinction
using the \citet{sf11} values as tabulated by the NASA/IPAC
Extragalactic Database (NED) for the appropriate ACS and WFC3/IR
bands. Specifically, the corrections were 0.403, 0.153, 0.109, and
0.063 mag in \gacs, \zacs, \jwfc, and \hwfc, respectively.  For our
error budget (Table~\ref{tab:errorbudget}), we included an uncertainty
of 10\% in the reddening corrections derived from these extinction
estimates \citep{sf11}.

\section{SBF and Color Measurements
\label{sec:sbfmeasurements}}

The SBF technique measures the intrinsic variance in a galaxy's
surface brightness distribution arising from statistical fluctuations
in the integrated stellar luminosity per pixel
\citep{ts88,jacoby92,cantiello03,raimondo05,cervino08}. For evolved
stellar populations, which predominate in early-type galaxies, stars
on the red giant branch (RGB) contribute most strongly to the
variance. The ratio of the variance to the mean surface brightness
scales inversely as the square of the distance; this ratio is
represented by the apparent SBF magnitude \mbar.  The distance is
obtained from a calibration of the corresponding absolute magnitude
\Mbar\ on the mean properties of the stellar population. At
space-based image resolution, the method is the most precise distance
indicator available for the general population of early-type galaxies
at ${\sim}10$ to 100 Mpc
\citep{biscardi08,blake09,blake10b,freedman10,jensen15}. The SBF
signal is particularly strong in the near-IR, where RGB stars are
brightest \citep{jensen03}, and the effects of dust extinction are
minimized.

The SBF analysis of the GO-15329 observations (labeled ``B'' for the
name of the PI) was performed by J. Jensen (JJ), while analysis of the
combined GO-14804 and GO-14771 data (labeled ``LT'' for the PIs) was
performed by M. Cantiello (MC), without communicating the results to
each other. Following the initial independent SBF analysis, to
cross-check the results, the IR images were exchanged and each
reduction procedure was then repeated for the other data set, again
without communicating the results.  The results were then shared with
J.~Blakeslee, who acted as a referee in comparing the two reductions.
This procedure yielded two independent SBF analyses for each of the
two independent data sets (B and LT) in both passbands.  Due to the
high degree of cross-checking inherent in this procedure, the
resulting SBF measurements are exceptionally robust.

Although the independent measurements were performed by the two
authors using different SBF analysis software, the basic SBF
measurement procedure is the same and has been described in detail
elsewhere
\citep{bva01,blake10b,cantiello05,cantiello07b,cantiello17,jensen03,jensen15}.
The first step was to determine the background level in the final
combined image. As a result of the limited field of view of WFC3/IR,
it was necessary to estimate the galaxy contribution to the background
by fitting an $r^{1/4}$ profile (\sersic\ model with $n{=}4$), which
provided a reasonable fit to the overall profile despite deviations
caused by the shell features \citep[][]{blanchard17,im17,palmese17}.
The range of background values over which acceptable fits were
obtained was used to estimate the uncertainty in the background, and
this uncertainty was propagated into the error budget for both the SBF
amplitude and \JH\ color (Table~\ref{tab:errorbudget}).

\begin{figure*}
\plotone{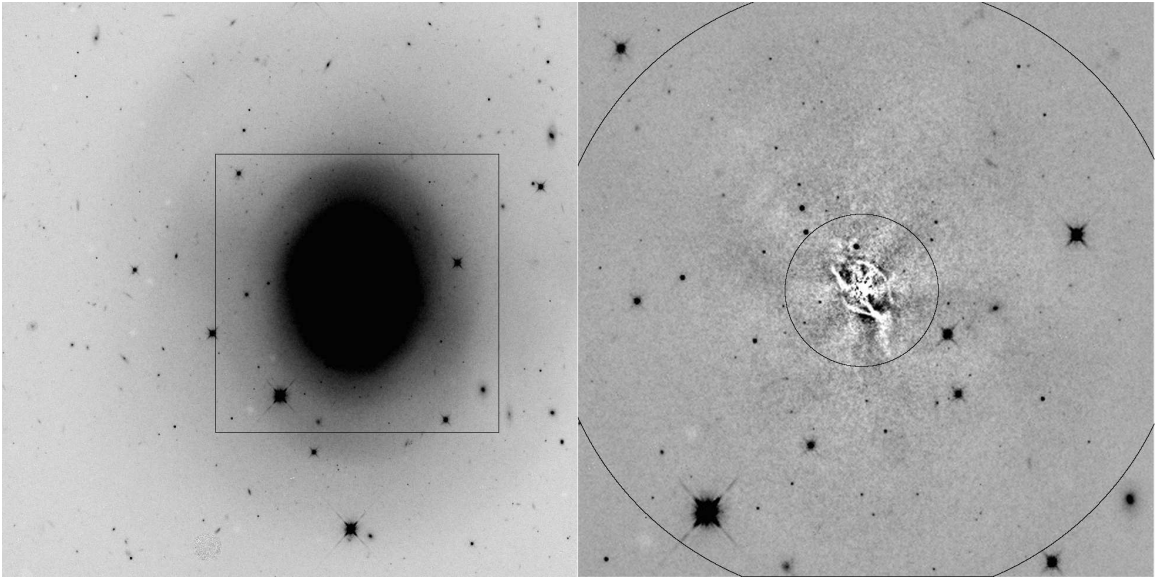}
\caption{{\it Left:} The full \jwfc\ image of NGC\,4993 (125 arcsec on
  a side; from GO-15329, PI: Berger) shown with a logarithmic scale to
  emphasize the faint outer shell structure (north is up, east is
  left). {\it Right:} The central square arcminute with the overall
  smooth light profile of the galaxy subtracted to reveal the narrow
  dust lanes near the center of the galaxy. The inner and outer limits
  of the radial region used for the SBF analysis are shown as circles.
\label{fig:imagefigure2}}
\end{figure*}

\begin{figure}
\plotone{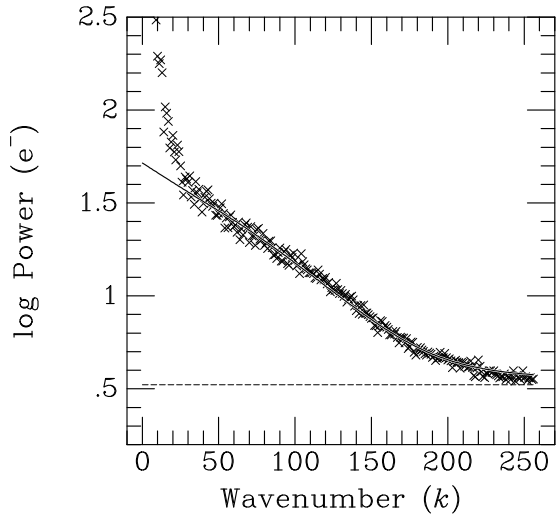}
\caption{The spatial power spectrum of the residual image of
  NGC\,4993, with dust lanes and other sources masked, fit as the sum
  of a scaled PSF power spectrum (solid line) and white noise
  component (dashed line). The upturn in the power spectrum at low
  wavenumbers $k$ occurs because of remaining large-scale features in
  the residual frame. The low $k$ range is excluded from the power
  spectrum fit \citep{cantiello17}. The apparent fluctuation magnitude
  \mbar\ is derived from the fitted power at $k{=}0$.
\label{fig:powerspectrum}}
\end{figure}

After background subtraction, the next step entailed modeling and
subtracting the two-dimensional galaxy light distribution and
large-scale residuals to obtain a clean residual image, as illustrated
in Figure~\ref{fig:imagefigure2}.  We then extracted bright stars to
create the PSF model. Because the SBF signal is convolved with the
PSF, an accurate determination of the PSF Fourier power spectrum was
essential. Contaminating sources such as foreground stars, background
galaxies, and especially globular clusters in the galaxy itself, were
identified using SExtractor \citep{bertin96} and masked in the
residual image. The SBF signal is the amplitude of the spatial power
spectrum of the masked residual image, normalized by the mean galaxy
surface brightness model, fitted with the normalized PSF power
spectrum (as shown in Figure~\ref{fig:powerspectrum}), and then
corrected for the residual power from undetected contaminating
sources.  The contribution from objects fainter than the limiting
detection threshold was estimated by fitting and extrapolating the
source luminosity function, as described in our previous
papers. Because these data are quite deep, and the SBF signal is very
strong, this correction was very small.  The corrected SBF amplitude
from the fitted spatial power spectrum
(Figure~\ref{fig:powerspectrum}) was then converted to the apparent
magnitude \mbar\ in the normal way and corrected for extinction.

All of these steps were followed independently by MC and JJ for
multiple circular annuli centered on NGC\,4993; the final measurements
were performed in an annulus extending from 8\farcs2 to 32\farcs8 from
the galaxy center (64 to 256 pixels, where the average pixel scale is
0\farcs128~pix$^{-1}$). Beyond this radius, the SBF and color
measurements were more strongly affected by uncertainties in the
background determination. Dust features are prominent at radii
interior to this annulus; the effect of dust is especially visible in
the optical ACS data, but is still visible at \jwfc\ in the right
panel of Figure~\ref{fig:imagefigure2}. The dust patches extending
beyond 8\farcs2 were masked using the multi-band color data.

The final \mbar\ measurements in each bandpass for each data set are
presented in Table~\ref{tab:distances}.  The tabulated error bars were
calculated by combining in quadrature the uncertainties in
\mbar\ arising from the background subtraction, power spectrum
fitting, PSF normalization, and the correction for contribution of
undetected point sources to the power spectrum
(Table~\ref{tab:errorbudget}).  All of these uncertainties are
discussed in detail in the references cited above.  The
\mbar\ measurements in Table~\ref{tab:distances} are used in the
following section to derive the distances.

As noted above, the distance estimation requires calibrating the
absolute SBF magnitude \Mbar\ based on the galaxy stellar population,
most commonly parameterized by the integrated galaxy color
\citep[e.g.][]{tonry97,bva01,blake09,jensen15,cantiello17}.  We
therefore used the ACS \gacs\ and \zacs\ images produced by the
standard STScI calibration pipeline to construct an optical color map
of the galaxy, transformed to the WFC3/IR distorted frame, and
measured the \gz\ color of the galaxy within the SBF analysis
region. Due to the larger ACS field of view, the sky backgrounds are
well determined, which allows the apparent color of this region of the
galaxy to be determined with an uncertainty of only ${\sim}0.01$ mag.
Including the estimated $10\%$ uncertainty on the Galactic reddening,
the corrected color measurement is $\gz=1.329\pm 0.027$ mag.

We also measured the extinction-corrected \JH\ color to obtain an
independent calibration of \Mbar. Due to the limited wavelength
coverage, this color index is not as constraining as \gz\ in
determining the absolute SBF magnitude. However, the reddening
correction is much smaller for \JH\ and adds an error of only
${\lesssim}\,0.005$ mag in quadrature, much less than for the optical
color. Thus, the additional information helped significantly to reduce
the uncertainty in \Mbar. As with the SBF analysis, the \JH\ color
measurements were performed independently by both JJ and MC from the B
and LT data sets, respectively. These measurements were averaged and
corrected for extinction, yielding $\JH = 0.259\pm 0.014$ mag.

\begin{deluxetable*}{cccccc}
\tablecaption{SBF Measurements\label{tab:distances}}
\tablewidth{0pt}
\tablehead{
\colhead{Dataset\tablenotemark{a} } &
\colhead{Filter} &
\colhead{\mbar} &
\colhead{\Mbar} &
\colhead{\mM} &
\colhead{Distance} \\
& & (AB mag) & (AB mag) & (mag) & (Mpc)
}
\startdata
\hline
\hline
\multicolumn{6}{c}{Using Eqn.\ 1 and 2, $\gz=1.329\pm 0.027$} \\
\hline
B  & F110W  & $30.041\pm0.056$ & $-3.040\pm0.077$ & $33.081 \pm 0.095$ & $41.3 \pm 1.9$ \\
LT &        & $29.999\pm0.068$ & $-3.040\pm0.077$ & $33.039 \pm 0.103$ & $40.5 \pm 1.9$ \\
B  & F160W  & $29.319\pm0.056$ & $-3.791\pm0.115$ & $33.110 \pm 0.128$ & $41.9 \pm 2.5$ \\
LT &        & $29.229\pm0.071$ & $-3.791\pm0.115$ & $33.020 \pm 0.135$ & $40.2 \pm 2.6$ \\
\hline
\hline
\multicolumn{6}{c}{Using Eqn.\ 3 and 4, $\JH=0.259\pm 0.014$}\\
\hline
B  & F110W  & $30.041\pm0.056$ & $-2.988\pm0.106$ & $33.029 \pm 0.120$ & $40.3 \pm 2.3$ \\
LT  &        & $29.999\pm0.068$ & $-2.988\pm0.106$ & $32.987 \pm 0.126$ & $39.6 \pm 2.3$ \\
B  & F160W  & $29.319\pm0.056$ & $-3.746\pm0.141$ & $33.065 \pm 0.152$ & $41.0 \pm 3.0$ \\
LT &        & $29.229\pm0.071$ & $-3.746\pm0.141$ & $32.975 \pm 0.158$ & $39.4 \pm 2.9$ \\
\hline
\hline
\multicolumn{6}{c}{Weighted average of both data sets and calibrations}\\
\hline                                           
BLT& F110W &$30.024\pm0.043$ & $-3.022\pm0.062$ & $33.046\pm0.076$ &$40.7\pm1.4$ \\
BLT& F160W &$29.284\pm0.044$ & $-3.773\pm0.089$ & $33.057\pm0.099$ &$40.9\pm1.9$ \\
\enddata
\tablenotetext{\rm a}{B = GO-15329 (PI E. Berger); LT = GO-14804 (PI A. Levan) + GO-14771 (PI N. Tanvir); BLT signifies the weighted average of the measurements from the B and LT data~sets.}
\end{deluxetable*}

\section{Distance Determination \label{sec:distance}}

To derive the distance modulus from the apparent SBF magnitude, \mbar,
we adopted a value for \Mbar\ from an empirical SBF calibration using
the galaxy \gz\ and \JH\ colors to correct for variations in stellar
population properties. The empirical SBF calibration of \Mbar\ used
here was derived from the distances to Virgo and Fornax cluster
galaxies, which are ultimately based on the Cepheid distance scale
\citep{tonry00,blake10b,jensen15}.  The \jwfc\ and \hwfc\ calibrations
from \citet{jensen15} are revised slightly from their published form
to take into account an improved characterization of the PSF model,
yielding a systematic offset of $+0.05\pm 0.02$ mag in \mbar\ for the
calibration sample compared to a much larger sample of SBF data
collected with \hst\ WFC3/IR after the \citet{jensen15} data were
collected, resulting in a much higher fidelity PSF measurement.  In
addition, the latest ACS photometric zero points imply that the
\gz\ color measurements from \citet{blake09} used for the calibration
were too red by +0.004 mag. With these updates, the calibrations (in
AB mag) are:

\begin{eqnarray}
 \overline{M}_{110}\;&=&\;-2.887 + 2.16\left[(g{-}z) - 1.4\right] \label{eq:jgz} \\
 \overline{M}_{160}\;&=&\;-3.640 + 2.13\left[(g{-}z) - 1.4\right] \label{eq:hgz} \\
 \overline{M}_{110}\;&=&\;-2.914 + 6.7\left[(J{-}H) - 0.27\right]  \label{eq:jjh}   \\
 \overline{M}_{160}\;&=&\;-3.668 + 7.1\left[(J{-}H) - 0.27\right]\,. \label{eq:hjh}
\end{eqnarray}

Following \citet{blake10b} and \citet{jensen15}, we adopt intrinsic
scatters of 0.05 and 0.10 mag for the $\overline{M}_{110}$ and
$\overline{M}_{160}$ calibrations, respectively. These estimates are
based on the observed scatter in the relations, corrected for the
effect of the measurement errors in both the color and SBF magnitudes
reported by \citet{jensen15}. As discussed in previous studies, the
observed scatter in \Mbar\ with integrated color is minimized at
wavelengths near ${\sim}1$ $\mu$m.

In Table~\ref{tab:distances} we report the eight individual distances
derived from the two independent measurements (B and LT) in each of
the two passbands, using the two different color calibrations. The
reported uncertainties in the \Mbar\ values include the intrinsic
scatter estimated from the calibration relations combined in
quadrature with errors propagated from the color measurements. We
present all of these estimates to illustrate good consistency;
however, these measurements are not all independent, and it would not
make sense to take a simple weighted average of all the distance
moduli.  Instead, we report in Table~\ref{tab:distances} the weighted
averages of the \mbar\ measurements from the two independent B and LT
data sets for each of the two passbands, combined with the weighted
average \Mbar\ values from the two color calibrations, to give the two
final distances derived from the \jwfc\ and \hwfc\ SBF measurements.
In each case, the largest contribution to the final error bar comes
from the adopted \Mbar, and the \Mbar\ estimates for the two bands are
based on the same color measurements.  Thus, we do not attempt to
average them; instead, we take the \jwfc\ result as our best
constraint on the NGC\,4993 distance and note that the \hwfc\ result
is nearly identical.

Finally, we note that the zero points of the calibration relations in
Equations~(\ref{eq:jgz}) to (\ref{eq:hjh}) are tied to the mean
distance modulus of $31.09\pm0.03\pm0.08$~mag to the Virgo cluster
based on 31 Virgo galaxies with distances measured in the ground-based
SBF survey of \citet{tonry01}. Here, the first error bar represents
the uncertainty in the mean, while the second represents the
systematic uncertainty in the tie of the SBF distances to the Cepheid
distance scale of \citet{freedman01} \citep[see the discussions
  by][]{blake10b,cantiello17}. Including an additional uncertainty of
0.06 mag for the Cepheid distance scale itself \citep{freedman10}, the
total systematic uncertainty in our \Mbar\ calibration is 0.10 mag.
Our final result for the SBF distance to NGC\,4993 is therefore
$(m{-}M)=33.05\pm0.08\pm0.10$ mag, corresponding to
$d=40.7\pm1.4\pm1.9$ Mpc (random and systematic errors, respectively).

\begin{deluxetable*}{lccl}
\tablecaption{SBF Distance Error Budget\label{tab:errorbudget}}
\tablewidth{0pt}
\tablehead{
\colhead{Uncertainty} &
\colhead{$\sigma_{110}$} &
\colhead{$\sigma_{160}$} &
\colhead{Source\tablenotemark{a}} \\
& \colhead{(mag)} & \colhead{(mag)} }
\startdata
\hline
\multicolumn{4}{c}{SBF Measurement Uncertainties} \\
\hline
Background & 0.01--0.015 & 0.005--0.01 & measured\\
PSF fit& 0.01--0.04 & 0.02--0.05 & measured\\
External source fit& 0.01--0.015 & 0.01--0.015 & measured\\
Spatial power spectrum fit& 0.05 & 0.05 & measured\\
\mbar\ total & 0.056--0.068 & 0.056--0.071 & added in quadrature\\
\hline
\multicolumn{4}{c}{Calibration Uncertainties}\\
\hline
PSF normalization & $0.02$ & 0.02 & comparison with J15 \\
\gz\ color correction & 0.027 & 0.027 &background, extinction (SF11) \\
\JH\ color correction & 0.014 & 0.014 &background, extinction (SF11) \\
Stellar population scatter & 0.05 & 0.10 & J15, B09 \\
\Mbar\ total & 0.077--0.106 & 0.115--0.141 & propagated and added in quadrature\\
SBF tie to Cepheid distance ZP & 0.10 & 0.10 & FM10, B10\\
\enddata
\tablenotetext{\rm a}{B09=\cite{blake09}; B10=\cite{blake10b}; FM10=\cite{freedman10}; J15=\cite{jensen15}; SF11=\cite{sf11}; }
\end{deluxetable*}

\section{Stellar Population of NGC\,4993 from SBF}
\label{sec:stellarpops}

The likely coalescence timescale for the GW170817 system can be
investigated using an estimate of the stellar population age of
NGC\,4993. \citet{blanchard17} reconstructed the star formation
history of the galaxy and found that half of the stellar mass was
assembled about 11 Gyr ago, with a negligible present-day star
formation rate of ${\lesssim}\,0.01$ M$_\odot$ yr$^{-1}$.
\citet{levan17} found that about 60\% of the stellar mass formed
$\gtrsim 5$ Gyr ago.  Both papers suggest that a merger occurred about
a Gyr ago based on the presence of dust lanes and shells, as well as
indications from the reconstructed star formation history.

\begin{figure}
\plotone{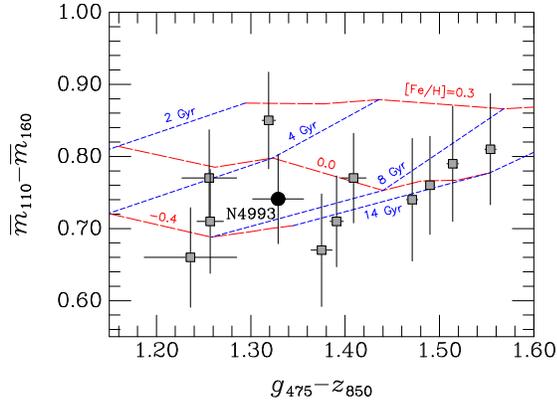}
\caption{{Surface brightness fluctuation colors are plotted
    vs.\ optical colors for NGC\,4993 (filled black circle) and a
    sample of elliptical and S0 galaxies from the Virgo and Fornax
    clusters, from \citet[][filled gray squares]{jensen15}.  SBF color
    is independent of distance, and therefore allows comparison with
    stellar population models to constrain the properties of the ages
    and metallicities of the galaxy's stars. The models have been
    shifted vertically by ${-}0.04$ mag, i.e., about one half of the
    intrinsic scatter of the models with respect to changes in some of
    the stellar population model ingredients. This was done to better
    match the predicted age of red massive galaxies in Virgo and
    Fornax with the accepted age of the Universe \citep{planck16}.
\label{fig:sbfcolors}}}
\end{figure}

The IR SBF signal arises almost entirely from RGB and AGB stars in
early-type galaxies, and variations in the SBF amplitude as a function
of radius or color can be used to probe the stellar population age and
metallicity of the dominant component of a galaxy \citep{jensen03}.
In Figure~\ref{fig:sbfcolors} we plot the distance-independent SBF
color \colorbar\ of NGC\,4993 versus the integrated color \gz,
together with previous measurements from \citet{jensen15} for 11
early-type galaxies in the Virgo and Fornax clusters, for comparison,
and with stellar population models. The SBF and integrated color
predictions shown in the figure are based on the SPoT single-age,
single-metallicity stellar population (SSP) models \citep{raimondo09}
originally presented by \citet{jensen15}, updated for this study using
a larger number of stars (stellar population mass
${\sim}\,2{\times}10^6~M_{\sun}$) and improved spectral libraries for
cooler stars.

NGC\,4993 has an SBF color that is very similar to the
\citet{jensen15} lenticular galaxies and lower-luminosity ellipticals
in Virgo and Fornax that have mean population ages of $\sim 6-10$ Gyr
and approximately solar metallicity.  The narrow wavelength interval
of the \JH\ SBF color does not allow us to place tight constraints on
the properties of the dominant stellar populations in the galaxy, but
the comparison with SSP models, shown in Figure\ \ref{fig:sbfcolors},
indicates that NGC\,4993 likewise has a luminosity-weighted stellar
population older than 6 Gyr with slightly sub-solar metallicity. This
is consistent with \citet{blanchard17} and \citet{levan17}, but using
a completely different technique that directly measures the properties
of the evolved giant branch stars.

The \gz\, and \JH\ colors show a modest gradient ($\Delta\gz\lesssim
0.1$ and $\Delta\JH\lesssim 0.03$ mag, respectively) with redder
colors near the galaxy center (excluding the dust lanes in the core),
similar to other early-type galaxies; the \gz\ color appears to
increase again at roughly 30\arcsec, apparently associated with a
shell feature. There appears to be no evidence of a trend in
fluctuation magnitude within the region used for the SBF measurement
(8\farcs2 to 32\farcs8 in radius). We conclude that, since NGC\,4993
shows signs of relatively recent merging (outer shells, central dust
lanes, and a change in the slope of the gradient in \gz), the
homogeneity of SBF measurements can be attributed either to a
well-mixed stellar population of the pre-merging systems, or to a
merging of galaxies with very similar stellar populations.

\section{Implications and Conclusions}
\label{sec:conclusions}

We have used \hst\ near-IR observations to measure the SBF distance to
NGC\,4993, leading to the most precise value available to date,
$d=40.7\,{\pm}\,1.4\,{\pm}\,1.9$ Mpc (random and systematic errors,
respectively).  This distance is consistent with the value
$d\,{=}\,43.8^{+2.9}_{-6.9}$ Mpc \citep{abbott17siren} estimated from
the GW data.  The SBF distance error of ${\sim\,}4$\% is much smaller
than the FP measurement uncertainty and significantly reduces the
uncertainty associated with the GW-derived distance
\citep{abbott17siren}.

While a single galaxy distance cannot place robust constraints on the
Hubble constant, we can check for consistency using our measured
distance and the recession velocity of the galaxy, and then use the
resulting $H_0$ to constrain the orbital inclination of the merging
BNS.  \citet{hjorth17} adopted a mean heliocentric velocity of $v_h =
2921\pm53$ \kms\ for the NGC\,4993 galaxy group. After transforming to
the CMB rest frame (a difference of $310$~\kms\ in this direction) and
correcting for an estimated peculiar velocity of $v_p=307\pm230$
\kms\ (the numerical similarity of $v_p$ to the projection along this
direction of the Sun's velocity in the CMB frame is coincidental),
they derive a Hubble-flow velocity\footnote{\citet{hjorth17} refer to
  $v_H$ as the ``cosmic velocity,'' which is not to be confused with
  the observed velocity in CMB frame, $v_{\rm CMB}$.} of $v_H = v_{\rm
  CMB} - v_p = 2924\pm 236$ \kms.  This value of $v_H$ agrees to
within 0.5\% of the independently estimated value from
\citet{guidorzi17}.  Taking the ratio, we find $H_0 = v_H/d =
71.9\,{\pm}\,7.1$ \kmsmpc, where the error bar includes both random
and systematic uncertainties.  Given the ${\sim}\,10\%$ uncertainty,
our inferred value of $H_0$ is consistent with both the Type~Ia
supernova measurements from SHoES (73.2 \kmsmpc; \citealt{riess16})
and the CMB measurement from Planck (67.7 \kmsmpc;
\citealt{planck16}).

For comparison, \citet{abbott17siren} inferred
$H_0=70.0^{+12.0}_{-8.0}$ \kmsmpc\ from a combination of the
GW-derived distance and an assumed $v_H{\,=\,}3017\,{\pm}\,166$ \kms.
The 3\% higher value of $v_H$ was based on a somewhat rougher estimate
of the mean observed velocity of the NGC\,4993 group (the adopted
peculiar velocity was nearly identical, although with a smaller
uncertainty, leading to the smaller quoted error bars).  Thus, to be
consistent with \citet{abbott17siren}, we need to multiply our value
of $H_0$ by 1.032. Applying this factor and comparing to the
  1-$\sigma$ curve in Figure~2 of that work, which presents the
degeneracy between $H_0$ and the binary orbital inclination, we find
that $i\,{\gtrsim}\,137$ deg. This is consistent with the 90\% upper
limit derived via the approach in \citet{mandel18}.

Finally, we emphasize that the distance measures to GW170817 are
estimated from two radically different and independent approaches, GWs
and SBF, and therefore the consistency is striking.

Looking to the future, we expect that (at the design sensitivity)
LIGO/Virgo will discover BNS mergers out to a few hundred Mpc, with
many events expected to occur within 300~Mpc.  Assuming that EM
counterparts will be detected for most of these mergers, the distances
to early-type host galaxies can be measured using the SBF technique
out to ${\sim}100$ Mpc with \hst, and possibly to ${\sim}300$ Mpc
using {\it James Webb Space Telescope} (based on estimates using the
available exposure time calculator).  Only Type Ia supernovae can
provide competitive distance measurements at these distances, but the
chances of observing a supernova in the same galaxy as a BNS merger
are unlikely.  In this paper we have demonstrated that SBF distance
measurements are a particularly compelling approach to breaking the
distance-inclination degeneracy of the GW data.

\acknowledgements This research has made use of the NASA/IPAC
Extragalactic Database (NED) which is operated by the Jet Propulsion
Laboratory, California Institute of Technology, under contract with
the National Aeronautics and Space Administration.  We acknowledge
funding from INAF project: Gravitational Wave Astronomy with the first
detections of aLIGO and aVIRGO experiments (PI E. Brocato).  MC,
GR and JPB acknowledge partial support from the PRIN INAF-2014
``EXCALIBURS: EXtragalactic distance scale CALIBration Using
first-Rank Standard candles'' project (PI G. Clementini).  The Berger
Time-Domain Group at Harvard is supported in part by the NSF through
grant AST-1714498, and by NASA through grants NNX15AE50G and
NNX16AC22G.  NRT acknowledges support from STFC consolidated grant
ST/N000757/1.  PDA and SC acknowledge support from ASI grant
I/004/11/3.  JH was supported by a VILLUM FONDEN Investigator grant
(project number 16599).  IM acknowledges STFC for partial support.
We thank the anonymous referee for a clear and unequivocal report.

\end{document}